\documentclass[twocolumn,aps,prl,10pt,amsmath,amssymb,nofootinbib,showpacs,superscriptaddress,floatfix]{revtex4-1}

\DeclareFontFamily{U}{rcjhbltx}{}
\DeclareFontShape{U}{rcjhbltx}{m}{n}{<->rcjhbltx}{}
\DeclareSymbolFont{hebrewletters}{U}{rcjhbltx}{m}{n}

\usepackage{graphicx}
\usepackage{color}
\usepackage[usenames,dvipsnames]{xcolor}
\usepackage[colorlinks=true,linkcolor=Red,citecolor=Green,linktoc=page]{hyperref}
\usepackage{multirow}
\usepackage{float}
\usepackage{flushend}
\usepackage{balance}
\usepackage[varg]{txfonts}
\usepackage{ulem}
\usepackage{fancyhdr}

\DeclareMathSymbol{\lamed}{\mathord}{hebrewletters}{108}

\begin{document}
\title{Berezinskii-Kosterlitz-Thouless quantum transition in 2 dimensions}

\begin{abstract}
\end{abstract}

\author{M.\,C.\,Diamantini}

\affiliation{NiPS Laboratory, INFN and Dipartimento di Fisica e Geologia, University of Perugia, via A. Pascoli, I-06100 Perugia, Italy}

\author{C.\,A.\,Trugenberger}

\affiliation{SwissScientific Technologies SA, rue du Rhone 59, CH-1204 Geneva, Switzerland}
\affiliation{Division of Science, New York University Abu Dhabi, Abu Dhabi, United Arab Emirates}

\author{V. M. Vinokur$\dagger$}
\affiliation{Sadly Dr. Vinokur passed away prior to the publication of this manuscript.}

\
\begin{abstract}
The Berezinskii-Kosterlitz-Thouless (BKT) transition is the prototype of a phase transition driven by the formation and interaction of topological defects in two-dimensional (2D) systems. In typical models these are vortices: above a transition temperature\,$T_{\rm BKT}$ vortices are free, below this transition temperature they get confined. In this work we extend the concept of BKT transition to quantum systems in two dimensions. In particular, we demonstrate that a zero-temperature quantum BKT phase transition, driven by a coupling constant can occur in 2D models governed by an effective gauge field theory with a diverging dielectric constant. One particular example is that of a compact U(1) gauge theory with a diverging dielectric constant, where the quantum BKT transition is induced by non-relativistic, purely 2D magnetic monopoles, which can be viewed also as electric vortices. These quantum BKT transitions have the same diverging exponent $z$ as the quantum Griffiths transition but have nothing to do with disorder. 
\end{abstract}
\maketitle


\section{Introduction}
The Berezinskii-Kosterlitz-Thouless (BKT) phase transition \cite{ber, kt1, kt2} is the paradigm of a thermal transition induced by topological defects. At high temperatures, vortices in a 2D statistical mechanics model, like the XY model, are free excitations and generate an exponential decay of correlation functions. Below a critical temperature $T_{\rm BKT}$ they are confined, with the consequence that correlation functions become algebraic. 

The original BKT transition is driven by the condensation of vortices as a function of temperature. Here we shall consider a different BKT transition, governed by a coupling constant, rather than temperature, but still in 2 space dimensions (2D). A continuous quantum phase transition in D dimensions is governed by the action of a field theory in (D+1) Euclidean dimensions (for a review see \cite{negele}). It is often stated that there cannot be a quantum ($T=0$) BKT transition driven by a coupling constant in a 2D model because the dimension of the Euclidean action would be 3, and would thus be too high (see, e.g. \cite{feigelman}). In this note we show that this is wrong. There are indeed 2D models that undergo a quantum BKT transition at $T=0$ and these involve gauge fields with a diverging dielectric constant. The mechanism is the same as in the original BKT transition but it is driven by a different class of topological defects.

A diverging dielectric constant is not an artificial construct; rather, it is crucially relevant for a real, measured quantum transition in 2D, as we now discuss. It was shown already in the 70s that the dielectric constant diverges at the metal-insulator transition \cite{efros}. When the temperature is lowered, the metal may turn into a superconductor. Then there is a corresponding quantum transition between this superconductor and an insulating state, which goes under the name superconductor-to-insulator (SIT) transition \cite{goldman1, haviland, hebard, fisher1, fisher2} (for a review see \cite{goldmanrev, feigelmanrev}), where the dielectric constant also diverges \cite{baturina, bakkali}. We shall discuss in detail the relevance of the quantum BKT transition due to this divergence for the SIT. In particular, the quantum BKT transition has the same diverging exponent $z$ as the quantum Griffiths transition (for a review \cite{vojta}) but is completely unrelated to disorder. What can be mistaken for a quantum Griffiths transition in thin superconducting films may be instead a quantum BKT transition. Note that the SIT transition to superinsulation has been recently shown to occur also in perfectly ordered systems, confirming that interactions are the dominant cause for the SIT \cite{bsi}. 

Intuitively, it is clear that a diverging dielectric constant $\varepsilon$ implies a dimension less, like infinite temperature. Recalling that the velocity of light in a medium is given by $v=1/\sqrt{\varepsilon \mu}$ (we use natural units $c=1$, $\hbar = 1$ and $\varepsilon_0=1$), with $\mu$ the magnetic permeability (which we will henceforth consider to be $\mu=1$ for simplicity), we see that the divergence of $\varepsilon$ implies a vanishing light velocity $v=0$. Therefore, nothing propagates in the medium, only static electric configurations matter, as follows from the 2D Maxwell equation
\begin{equation}
v^2 \epsilon^{ij} \partial_j B= \partial_t E^i \ .
\label{max}
\end{equation}
Since time-dependent electric fields are the only source of magnetic fields in vacuum, these vanish in the limit of diverging dielectric constant and the Lagrangian
becomes positive definite. As a consequence, the Euclidean quantum partition function is determined by an effective 2D Hamiltonian of a statistical mechanics model of static electric configurations with a quantum coupling of canonical dimension [mass] playing the role of ``temperature". 

To show formally how this dimensional reduction works, let us consider the usual electromagnetic Hamiltonian (in (2+1) dimensions) in a medium with a dielectric constant $\varepsilon$,
\begin{equation} 
H = {1\over 2 f^2} \int d^2 x \  {\bf D} {\bf E} + B^2 = {1\over 2 f^2} \int d^2 x \ \varepsilon {\bf E}^2 + B^2 \ ,
\label{hameps}
\end{equation}
where $f^2$ is the usual massive coupling of (2+1)-dimensional electromagnetism which, in the case of applications to thin films, is given by $f^2= e^2/d$ with $d$ the film thickness and $e$ the electron charge. The corresponding Lagrangian is 
\begin{equation} 
L = {1\over 2v^2 f^2} \int d^2 x \  {\bf E}^2 - v^2 B^2 \ ,
\label{lageps}
\end{equation}
from where it is evident that magnetic fields become irrelevant in the limit $v\to 0$, $f^2 \to \infty$ with $v^2 f^2 = \phi^2= {\rm constant}$, which corresponds to the strong coupling regime of the model.  In this limit the Lagrangian becomes
\begin{equation} 
L = {1\over 2\phi^2} \int d^2 x \  {\bf E}^2  \ ,
\label{laglim}
\end{equation}
which is positive definite and involves only static electric field configurations. 

The Euclidean quantum partition function $Z = \int {\cal D} {\bf A} \ {\rm exp}(-S)$ involves the Euclidean action. At a finite temperature $T = 1/\beta$ ($k_B=1$ for simplicity) this is 
\begin{equation} 
S={1\over 2\phi^2} \int_0^{\beta} dt \int d^2{\bf x} \ {\bf E}^2 \ ,
\label{add1}
\end{equation}
with periodic boundary conditions in time $t$ for the bosonic electromagnetic fields. We now change variables to $\tau = vt$. Recalling that the electric field is  ${\bf E} = -\nabla A_0 -\partial_t {\bf A}$ and that $A_0$ scales like $1/t$ we obtain 
\begin{equation}
S = {1\over 2\phi^2} v \int_0^{\beta v} d\tau \int d^2 x \ {\bf E}^2 \ ,
\label{ac1}
\end{equation}
where now ${\bf E} = -\nabla (A_0/v) -\partial_\tau {\bf A}$. 
Finally, since electric fields are static, we can evaluate the integral over $\tau$ to obtain
\begin{equation}
S = {1\over 2\phi^2} \beta v^2 \int d^2 x \  {\bf E}^2  \ .
\label{ac2}
\end{equation}
We can now take the limit of zero temperature $T\to 0 \  (\beta \to \infty)$ simultaneously with $v\to 0$ so that $\beta v^2$ is a fixed constant $b$. This is the appropriate limit to describe a quantum transition with diverging dielectric constant, like the SIT. By redefining rescaled electric fields as $\tilde {\bf E} = {\bf E}/\phi$, we obtain an effective statistical mechanics model with a Hamiltonian $\tilde H =(1/2) \int d^2 x \ \tilde E^2$ and an effective temperature $1/b$. Since the real temperature is zero, this describes only quantum fluctuations. 

This illustrates the dimensional reduction due to the limit $\varepsilon \to \infty$. Unfortunately, there are no static electric field configurations inducing a phase transition in this version of electrodynamics. As is well-known since the famed work of Polyakov \cite{polyakov, polyakovbook} in the 70s, however, there are two versions of electrodynamics. One is non-compact, with gauge group ${\mathbb R}$, while the other is compact, with gauge group $U(1)$. As we now explain, in the framework of applications to granular condensed matter systems, like superconducting films near the SIT, it is always the compact version which is the relevant one. Compact electrodynamics in (2+1) dimensions is a cutoff theory. There are two possible ways to implement the cutoff: one is to obtain the model by spontaneous symmetry breaking from a non-Abelian gauge theory, the other one is to formulate the model on a lattice. The lattice formulation automatically arises in granular media. And in this case things change considerably, there are indeed static  configurations inducing a quantum BKT transition.

\section{The Keldysh potential}
Thin films are typically modelled as systems in which the matter degrees of freedom are constrained to live on a 2D plane, while electromagnetic fields can space out in 3D. But do they really do so? Let us consider the experimentally relevant configuration of a horizontal thin film of large dielectric constant $\varepsilon$. In this geometric arrangement, there is necessarily a half-space above the film and one below the film and these can be considered as filled by media, which can of course also be the vacuum. So, above the film there is a medium of dielectric constant $\varepsilon_1$ and below the film there is a medium of dielectric constant $\varepsilon_2$. If the film is suspended in vacuum, we have $\varepsilon_1=\varepsilon_2 = 1$. If the film is deposited on a substrate, only $\varepsilon_1=1$. We are interested in the case in which $\varepsilon_1 \ll \varepsilon $ and $\varepsilon_2 \ll \varepsilon$. This problem has been considered and solved in \cite{chaplik, keldysh} (for a review see \cite{baturina}) and leads to the electric Keldish potential, which is a purely 2D logarithmic Coulomb potential up to an electric screening length $\lambda_s= d \varepsilon/(\varepsilon_1+\varepsilon_2)$, with $d$ the film thickness. 

This result can also be obtained with a much simpler calculation in the symmetric situation where the media above and below the film are the same. In this case, the behaviour of the electric field at the upper and lower boundaries of the film must be the same by symmetry: both normal components vanish due to the much higher dielectric constant in the film than outside and therefore the potential must also be the same since a normal component inside the film has nowhere to end. This implies periodic boundary conditions. With periodic boundary conditions, the Coulomb potential $G$ among two charges $e$ in a film of thickness $d$ along the z-axis is determined by the equation
\begin{equation}
\nabla^2 G\left( |{\bf x}|, z, d \right)  = e^2 \delta^2 \left( {\bf x} \right) \sum_{n=-\infty}^{+\infty} \delta(z-nd) \ ,
\label{cpot1}
\end{equation}

The periodic delta function represents an infinite series of couples of mirror charges on the z-axis outside the film, whose effect is to mimic the periodic boundary conditions. We now recall that an electric dipole of length $\ell$ in a medium with dielectric constant $\varepsilon$ depends on the effective distance $\ell/\varepsilon$. Therefore, to maintain periodic boundary conditions with the same thickness $d$ but with a dielectric constant $\varepsilon$ inside the film we must place the mirror charges at distances $n \varepsilon d$,
\begin{equation}
\nabla^2 G\left( |{\bf x}|, z, d \right)  = e^2 \delta^2 \left( {\bf x} \right) \sum_{n=-\infty}^{+\infty} \delta(z-n\varepsilon d) \ .
\label{cpot2}
\end{equation}
The solution to this equation can be written in the form
\begin{equation}
G\left( |{\bf x}|, z, d \right) = {e^2\over 4\pi \varepsilon d} \sum_{n=-\infty}^{+\infty} \int_0^{\infty} dt {1\over \sqrt{1+t^2}} {\rm e}^{i {2\pi n\over \varepsilon d} \left( t|{\bf x}|-z \right)} \ . 
\label{cpot3}
\end{equation}
In the regime $|{\bf x}| \gg \varepsilon d$ the prefactor of $t$ in the phase is very large even for $n=\pm 1$. Therefore the phase factor oscillates wildly and the whole expression is suppressed by the integral over $t$. In the opposite regime $|{\bf x}| \ll \varepsilon d$ the phase again oscillates strongly for high values of $n$. In this regime, the dominant contribution that survives comes from the first dipole $n=\pm 1$. In this regime we can thus keep only the first dipole $n=\pm 1$ and neglect $z$ in the ensuing cosine function, which gives
\begin{equation}
G\left( |{\bf x}|, d \right) =  {e^2\over 4\pi \varepsilon d} \int_0^{\infty} dt {1\over \sqrt{1+t^2}} {\rm cos} {2\pi \over \varepsilon d}  t|{\bf x}| =
{e^2\over 4\pi \varepsilon d} K_0 \left( 2\pi {|{\bf x}| \over \varepsilon d} \right) \ ,
\label{cpot4}
\end{equation}
where $K_0$ is the zeroth-order modified Bessel function of the second kind. As expected, both (square) charges and 2D distances are renormalized by a factor $1/\varepsilon$. This result becomes exact in the limit $d\to 0$, $\varepsilon \to \infty$ so that $\varepsilon d = \lambda_s$ is finite, in which case the Coulomb potential becomes a pure 2D one,
\begin{equation}
G\left( |{\bf x}|, d \right) = - {e^2\over 4\pi \lambda_s} {\rm ln} \left( {|{\bf x}| \over \lambda_s }\right) + {\rm constant} \ ,
\label{copt5}
\end{equation}
up to a screening length $\lambda_s$. Up to a numerical constant O(1) this is the Keldish result. The Keldish potential implies that, for very large dielectric constant $\varepsilon$, the electric screening length in very thin films can become typically larger than the lateral dimension $L$ of the sample, $L < \lambda_s$. In films where this happens, electromagnetism is entirely 2D: the lines of electric field do not exit the film in the orthogonal direction. 

This has a momentous consequence if the material in the film is a superconductor. Indeed, the time-dependent Ginzburg-Landau theory becomes a super-renormalizable, purely 2D theory and, when the strong electric fluctuations are taken into account, is therefore plagued by infrared divergences. These can be cured by an expansion in terms of ${\rm ln} (e^2 L/d )$ \cite{jackiw}, which means the theory becomes non-perturbative in systems larger than their thickness. When decreasing the thickness, a film of a superconductor with a large normal-state dielectric constant breaks thus up in many condensate islands of the typical lateral size $d$, each characterized by its own phase, forming an emergent Josephson junction array (JJA) \cite{super2D, typeIII, typeIIIvort}. This has been verified for such materials as TiN \cite{sacepe1}, NbTiN \cite{nbtin} and InO \cite{granular, kapitulnik}. It becomes now clear that, in these granular materials, it is the compact version of electrodynamics which is the relevant one. Each condensate island is characterized by its own independent phase: since this is an angular variable, the gauge group is the compact group $U(1)$. The compactness of the gauge group implies the presence of topological excitations, absent in the non-compact version of electrodynamics. We must thus amend the above computation of the effects of a large dielectric constant to take these topological excitations into account.

\section{Compact QED with a diverging dielectric constant}
Thin superconducting films of high-$\varepsilon$ materials can exist in three possible quantum ground states: type-III superconductors \cite{typeIII}, Bose metals \cite{dst} (for a review see \cite{boserev} and superinsulators \cite{dst, vinokurnature, baturina, dtv} (for a review see \cite{enc}). In this last state, the induced electromagnetic action is compact electromagnetism, with the cutoff lattice structure automatically provided by the emergent JJA granularity. 

Let us start from the lattice version of the action (\ref{add1}),
\begin{equation}
S= {1\over 2v^2 f^2} N\ell_0  \sum_{\{{\bf k}, i=1,2\}} \ell^2 \  E_i^2\ ,
\label{rev1}
\end{equation}
where $\ell$ is the spacing on a 2D lattice with sites labelled by integers {\bf k }, representing the typical distance of the condensate islands and $\beta = N\ell_0= N\ell /v$, with N the number of lattice spacings $\ell_0$ in the Euclidean time direction. Compactness of the gauge group means that the gauge fields $A_{\mu}$ are not real variables but, rather are angular variables defined in the interval $[-\pi/\ell_0, +\pi/\ell_0]$ for $A_0$ and $[-\pi/\ell, +\pi/\ell]$ for $A_i$. This entails that the action must remain unchanged when the (dimensionless) electric fields $\ell_0 \ell E_i$, which are finite differences of gauge fields, are shifted by an integer multiple of $2\pi$. This, in turn, can be easily achieved in the usual Villain-type formulation by introducing dynamical integer link variables $K_i$ and summing over them in the partition function. The action is modified to
\begin{equation}
S= {1\over 2v^2 f^2} N\ell_0 \sum_{\{{\bf k}, i=1,2\}} \ell^2 \left( E_i -{2\pi \over \ell_0 \ell} K_i \right)^2\ .
\label{compac1}
\end{equation}

Electric and magnetic fields can be represented compactly in terms of the Maxwell tensor $F_{\mu \nu}$, where Greek subscripts denote space-time indices, as opposed to Latin subscripts, which indicate purely spatial indices. Moreover, we can also introduce the dual electromagnetic tensor $F_\mu =(1/2)  \epsilon_{\mu \alpha \nu} F_{\alpha \nu}$, where $\epsilon_{\mu \alpha \nu}$ is the totally antisymmetric tensor and a sum over the same Greek indices is understood. In (2+1) dimensions, this is actually a three-vector, rather than a tensor and electric fields are given by $E_i = \epsilon_{ij}F_j$. In terms of these variables, the compact action takes the form 
\begin{equation}
S= {1\over 2v^2 f^2} N\ell_0  \sum_{\{{\bf k}, i=1,2\}} \ell^2 \left( F_i -{2\pi \over \ell_0 \ell} M_i \right)^2\ ,
\label{compac1}
\end{equation}
where $M_i =- \epsilon_{ij} K_j$. These dynamical integer link variables $M_i$ are the topological excitations which implement the compactness of the gauge model \cite{polyakov, polyakovbook}.  When viewed from the 3D Euclidean space point of view, $F_{\mu}$ represents an effective ``magnetic field", since $A_0$ plays exactly the same role as the component $A_3$ in the usual 3D electromagnetic action. Therefore, the components $F_i$ can be viewed as the corresponding spatial components of this effective 3D magnetic field on the film plane. The correct value of the dimensionful coupling constant $f^2$ can be read off the Coulomb potential derived in (\ref{copt5}), $f^2 = e^2/4\pi d$. Using this value, the action can be finally rewritten as 
\begin{equation}
S=  {1\over \pi} g\eta   \sum_{\{{\bf k}, i=1,2\}} \left( \ell_0 \ell F_i -2\pi M_i \right)^2\ ,
\label{compac2}
\end{equation}
where
\begin{eqnarray}
g &&= {2\pi^2 d  \over e^2 v \lambda_L } Nv\ ,
\nonumber \\
\eta &&= {v\lambda_L \over \ell} \ ,
\label{couplings}
\end{eqnarray}
with $\lambda_L$ the London penetration of the superconducting material. Note that this quantity falls out of the product $g\eta$. We have introduced it here to make contact with the two dimensionless coupling constants $g$ and $\eta$ that drive the SIT \cite{dst, bm}. 

Contrary to the non-compact case described earlier, compactness requires topological excitations encoded in $M_i$ which potentially drive a phase transition. Let us derive what exactly these topological excitations are. To do so we follow the standard treatment and decompose the integers $M_i$ in their transverse and longitudinal components. The most general such decomposition is \cite{polyakov, polyakovbook} 
\begin{equation}
M_i = \epsilon_{ij} \left( \Delta_j (N_0+\xi_0) - \Delta_0 (N_j+\xi_j) \right) + \Delta_i \chi \ ,
\label{dec}
\end{equation}
with $N_{0,i}$ integers and $\xi_{0,i}$ and $\chi$ real. Here, $\Delta_{0,i}$ denote finite differences on the lattice (note that a proper treatment of gauge invariance requires introducing slightly different lattice operators but this is not of importance here). The important point is that the transverse and longitudinal components of the integers $M_i$ need not be separately integers, they must only sum to integers. However, when applying a finite difference operator on $M_i$, the transverse component falls out. Since differences of integers are still integers, 
the longitudinal components are restricted by the equation
\begin{equation}
\Delta_i M_i = \nabla^2 \chi = m 
\label{mon1}
\end{equation}
with $m \in {\mathbb Z}$ and $\nabla^2$ the 2D finite difference Laplacian. Since $M_i$ have the meaning of integer magnetic field lines along the film in 3D Euclidean space-time, the integers $m$, the sources of these magnetic field lines, are non-relativistic magnetic monopoles, emanating their magnetic field only along the film plane. These topological configurations represent the extreme non-relativistic limit of Polyakov's famous magnetic monopoles\cite{polyakov, polyakovbook}. Note that these are magnetic monopoles only from the point of view of Euclidean 3D space-time. In Minkowski (2+1)-dimensional space-time they represent tunnelling events (instantons) in which a vortex made by a non-trivial circulation of the phases on neighbouring islands appears or disappears from the ground state. These tunnelling events are accompanied by closed, integer electric field loops $E^i = \epsilon^{ij}M_j$ around them. Since these electric field form loops, no electric charge is involved in these topological excitations, they are purely magnetic. Actually, at the transition point, where $v=0$, the tunnelling time becomes infinite, so that the topological excitations are ``frozen instantons", corresponding to static electric vortices. The plasma of magnetic monopoles generates a linear potential, representing a string of electric field between probe electric charges in the film \cite{dtv, enc}. Due to this string, charges in the film are confined, exactly as quarks in mesons, and cannot be separated on distances larger than the string dimension, leading to the infinite resistance (even at finite temperatures) of superinsulators. This is the phenomenon of charge confinement. The linear potential leading to charge confinement has been directly measured in experiment \cite{relaxation}. 

Let us now investigate how these magnetic monopole topological excitations appear in the compact electromagnetic action (\ref{compac2}). If we recall the definition of the fields $F_i$ in terms of gauge potentials $A_0$ and $A_i$ introduced above, we recognize that the transverse components of the integers $M_i$ appear exactly as additional gauge fields. In particular, 
the integers $N_{0,i}$, over which we sum in the partition function, can be used to shift the integration domain of the original gauge fields $A_0$ and $A_i$ from a circle to the entire real line. Then, also the real $\xi_{0,i}$ can be reabsorbed into the definition of the gauge fields, which are now real variables anyway. This gives back the action $S_0$ of the non-compact model considered above, only the longitudinal components of $M_i$ encode the topological excitations. At this point, since $F_i$ is purely transverse and the remaining $M_i$ is purely longitudinal, the cross-product term $F_i M_i$ in the action vanishes. The residual additional term $M_i^2$ can be expressed in terms of the monopole excitations by using (\ref{mon1}).
The total action decomposes thus as $S=S_0 + S_{\rm top}$, with the additional piece 
\begin{equation} 
S_{\rm top} = 4\pi g \eta  \sum_{\{{\bf k}\}} m {1\over -\nabla^2} m  \ ,
\label{topac}
\end{equation}
due to the gauge group compactness. The Euclidean 3D partition function of the compact gauge theory in the limit of large dielectric constant corresponds formally to the 2D statistical mechanics model of a Coulomb gas (for a review see \cite{minnhagen}) with coupling $J$ and temperature $T$ such $J/T = g\eta$. This undergoes a BKT phase transition, which corresponds, in the present case, to the logarithmic confinement of magnetic monopoles. At this transition, electric charges are correspondingly liberated and superinsulation is destroyed. At the transition, the correlation length diverges, which means, as usual, that we can take the continuum limit by removing the ultraviolet cutoff, in this case $\ell$, and therefore also $d$, which is of the same order of magnitude. The quantum BKT transition corresponds thus to the simultaneous limits $d \to 0$, $\ell \to 0$, $N \to \infty$ and $v\to 0$ so that $d/v$, $\ell/v$ and $Nv$ are finite, in which case the coupling takes a finite value $(g\eta)_{\rm cr}$, which sets the critical temperature $J/T_{\rm cr}$ of the corresponding Coulomb gas. It turns out that the BKT critical value is actually $(g\eta)_{\rm cr} = 1$ with $\eta > 1$ \cite{dst, dtv}. The transition in the compact gauge theory is thus a zero-temperature quantum BKT transition with diverging dielectric constant. 

We would like to conclude this section with a caveat and a note. First of all, the existence of the 2D quantum BKT transition relies on the particular limit described above and it is not guaranteed that all systems will end up in this regime. Our point is to prove that such a quantum BKT transition in 2D {\it can} exist, which is totally new and unexpected. Secondly, its realization depends on an underlying structure and the effective electromagnetic action being the compact version. This is not exotic. As we have seen, the condensate of thin superconducting films near the SIT typically decomposes into grains that provide the necessary structure and the superinsulating state of such films does indeed realize the compact version of QED.

\section{Vortex transition and the SIT}
Another example of a quantum BKT transition in 2D is the dual transition out of a superconductor. In this case, the relevant effective field theory is an emergent gauge theory for vortices \cite{typeIII, typeIIIvort} with Euclidean discrete action
\begin{eqnarray}
S &&= {1\over 2v^2 k^2} \sum_t \ell_0 \sum_{\{{\bf k}, i=1,2\}} \ell^2 \ \left( {\bf e}^2 + v^2 b^2 \right) 
\nonumber \\
&&+ i  \sum_t  \sum_{\{{\bf k}, i=1,2\}} \ \ell_0 a_0 \varphi_0 + \ell a_i \varphi_i \ ,
\label{gaugevort}
\end{eqnarray}
where $a_{\mu}$ is the emergent gauge field, ${\bf e}$ and $b$ the corresponding emergent electric and magnetic fields, respectively, $\varphi^{\mu}$ the integer vortex three-current  
and $k^2$ the relevant mass coupling constant. In the limit of high effective dielectric constant there are no currents and again, only static emergent electric configurations matter. We can repeat verbatim the steps above to obtain the action
\begin{equation}
S = {1\over 2 v^2 k^2 \ell} vN \sum_{\{{\bf k}, i=1,2\}} \left( \ell_0 \ell {\bf e}^2 \right) + i N \sum_{\{{\bf k}, i=1,2\}} \ \ell_0 a_0 \varphi_0 \ ,
\label{gaugevorthigh}
\end{equation}
This defines again an effective 2D statistical mechanics model. Upon integrating out the emergent gauge field this becomes once more a Coulomb gas, this time for vortices $\Phi = \varphi_0$, 
\begin{equation} 
S_{\rm top} = {Nv \over 2k^2 \ell}  \sum_{\{{\bf k}\}} \Phi {1\over -\nabla^2} \Phi  \ ,
\label{vortrans}
\end{equation}
The corresponding BKT transition survives at zero temperature, $N\to \infty$, if the dielectric constant diverges, $v\to 0$, so that $Nv \to {\rm constant}$. Using the relevant coupling constant for the SIT one obtains \cite{dst}
\begin{equation} 
S_{\rm top} = 4\pi {g \over \eta}  \sum_{\{{\bf k}\}} \Phi {1\over -\nabla^2} \Phi  \ ,
\label{vortrans}
\end{equation}
In this case, the vortex liberation and the ensuing destruction of superconductivity happens at the critical value $(g/\eta)_{\rm cr} = 1$ for $\eta >1$ \cite{dst}. Again, this is a purely quantum BKT transition at $T=0$, associated with a diverging dielectric constant. 

For $\eta > 1$, we have thus two quantum BKT transitions at $g=\eta$ and at $g=1/\eta$. For high values of $g > \eta$ we have a superconductor, for very low values of $g< 1/\eta$ we have a superinsulator. At both transitions the effective dielectric constant diverges. In between there is a Bose metal phase, a topological state in which both charges and vortices are out of condensate but are frozen by statistical interactions \cite{dst, bm} (for a review see \cite{boserev}). For $\eta < 1$, instead,  there is a domain $\eta  < g <1/\eta$ in which superinsulation can coexist with superconductivity, a superinsulator being the most stable state for $g<1$ while a superconductor is favoured for $g>1$. This is essentially a ``first-order quantum transition" in which the changeover from superconductivity to superinsulation is discontinuous \cite{dst, bm}. The two lines of phase transitions between superconductor and Bose metal and between Bose metal and superinsulator and the interactions between charges in these three phases have been derived analytically in \cite{dst, bm, dtv}. Here the focus is on their BKT universality class and the generic mechanism by which quantum BKT transitions in 2D may take place.

\section{Diverging dynamical exponent $z$ with no disorder}
As mentioned above, a continuous quantum phase transition in D dimensions is governed by the action of a field theory in (D+1) dimensions. This field theory must not necessarily be relativistic. In general, it is characterized  by two different scaling exponents for space and time variables (for a recent review see \cite{baiguera})
\begin{eqnarray}
{\bf x} && \to \lambda {\bf x} \ ,
\nonumber \\
t && \to \lambda^z t \ ,
\label{scalings}
\end{eqnarray} 
where the case $z=1$ corresponds to Lorentz invariant theories, while any $z>1$ encodes deviations from relativistic theories, the best known one being perhaps $z=2$ for Schr\"odinger field theory. Correspondingly, the continuous quantum transition is characterized by two different critical exponents $\nu$ and $z\nu$ for the spatial and (Euclidean) time directions, defined by the respective divergences of the correlations length $\xi$ and $\xi_t$ 
\begin{eqnarray}
\xi && \propto \left( \alpha - \alpha_{\rm c} \right)^{-\nu} \ ,
\nonumber \\
\xi_t && \propto \left( \alpha - \alpha_{\rm c} \right)^{-z\nu} \ ,
\label{exponents}
\end{eqnarray} 
where $\alpha$ is the relevant quantum coupling and $\alpha_{\rm c}$ its critical value. The case $z=2$ is known as a quantum Lifshitz transition. 

Of particular interest is the extreme case $z\to \infty$, which is normally associated with a quantum Griffiths transition (for a review see \cite{vojta}) corresponding to ``infinite" disorder. 
Our result shows that this is not necessarily the case. As we have shown, the extreme non-relativistic case corresponding to $z\to \infty$ is realized as well for quantum BKT transitions due to a diverging dielectric constant, in which case it has nothing to do with disorder and can occur also in perfectly ordered JJA. This is what has induced some authors \cite{xing, zhang, jing, yadav} to associate the superconductor to Bose metal transition with a quantum Griffiths transition due to extreme disorder. This is already an improbable scenario because it would entail a discontinuity of the quantum transition at $T=0$ with respect to $T>0$, where it is known to be a BKT transition. Given that, in these thin superconducting films, the dielectric constant is known to diverge at this transition, the alternative scenario of a quantum BKT transition independent of disorder \cite{bm} should be highly favoured. Of course, in this case also the critical exponent $\nu$ diverges. In other words, a diverging $z$ exponent can be indicative of both a quantum Griffiths transition or a 2D quantum BKT transition. However, the Griffiths scenario must be discarded in favour of the 2D quantum BKT transition when the $z$ divergence occurs in a perfectly ordered system.

In conclusion, a zero-temperature quantum BKT transition in 2D is possible if the dielectric constant diverges. Such a quantum BKT transition is realized in thin superconducting films at the superinsulator to Bose metal and superconductor to Bose metal transitions. At these transitions the critical exponent $z$ diverges even if they have nothing to do with disorder, having been observed also in perfectly ordered systems \cite{bsi}. It is thus highly likely that the experimental observations of a diverging critical exponent $z$ in thin superconducting film transitions is a confirmation of a quantum BKT transition.

\end{document}